\pdfoutput=1
\RequirePackage{amsmath}
\documentclass{llncs}
\usepackage{amsfonts}
\usepackage{amssymb}
\usepackage{graphicx} 
\usepackage[hyphens]{url}
\usepackage{hyperref}
\usepackage[utf8]{inputenc}
\usepackage[english]{babel}
\usepackage{booktabs}
\usepackage{numprint}
\usepackage[amsmath,thref,thmmarks]{ntheorem}
\usepackage{listings}
\usepackage{tabularx}
\usepackage{float}
\usepackage{upquote} 
\usepackage{eurosym} 
\usepackage[mathletters]{ucs} 
\usepackage{fancyvrb} 
\usepackage{tikz}
\usetikzlibrary{shapes.geometric,arrows}
\usepackage{xcolor}
\usepackage[draft]{todonotes}  

\definecolor{incolor}{rgb}{0.0, 0.0, 0.5}
\definecolor{outcolor}{rgb}{0.545, 0.0, 0.0}

\newcommand*{\figuretitle}[1]{%
\begin{flushleft}
 {\textbf{#1}
 \par\medskip}
\end{flushleft}
}

\begin{document}
\pagestyle{headings} 

\mainmatter       

\title{Intertemporal Connections Between Query Suggestions and Search Engine Results for Politics Related Queries}

\author{
Malte Bonart\inst{1} 
\and 
Philipp Schaer\inst{1}
}

\institute{
Technische Hochschule Köln, 50678 Cologne, Germany\\
\email{firstname.lastname@th-koeln.de}\\
}

\maketitle
\noindent

\section{Introduction}

This paper deals with the combined analysis of two data sources: First the AlgorithmWatch data-set on search results for political queries during the German Federal Elections in 2017 \cite{zweig2018} and second data from a long-term data crawl of search engine query suggestions. This enables us to examine result lists and query suggestions for the same set of politics related queries over a 2 month period. 

Query suggestions comprise a set of interactive techniques to assist users with the formulation of their query. As Kelly et al. \cite{kelly_comparison_2009} point out they seem particularly important in cases where users are searching for topics on which they have little knowledge or familiarity and in which they are in a so called ``anomalous state of knowledge'' \cite{belkin_ask_1982}. In the context of political opinion formation little is known about the influence of query suggestions in driving the users information behaviour. Regarding search results, the ``search engine manipulation effect'' was measured in several experimental setups \cite{epstein_search_2015}. It claims that biased search rankings can shift the voting preferences of undecided voters. However, the paper is criticized for presenting exaggerated effect sizes \cite{zweig2017}. 

In this paper we (1) present a data set with query suggestions crawled from different web search engines to complement the AlgorithmWatch data. As a first approach we (2) compare the aggregated \emph{ranking stability} of search results and query suggestions for $16$ search queries over a $2$ month observation period. Here, stability refers to the degree of similarity of successive lists over time. We suspect that external events both influence the composition of query suggestions and search results and the goal is to better understand how these two search engine features interact with each other through the search behaviour of the users. \footnote{E.g. an external event such as the marriage of a politician, may change his or her result set since websites dealing with this event become more prominent. In addition, search engine users will possibly also search for the event in connection with the politician's name which influences the composition of the query suggestions.} In a last step, we (3) shortly discuss further research topics. 

\section{Data Sets}

Our data-set consists of query suggestions (e.g. autocompletions) to complement the search engine results from the AlgorithmWatch data. For most queries an exact match can be obtained.\footnote{\label{fn:missing}The term ``cdu" is missing in our data-set. The search-terms ``Bündniss90/Die Grünen" and ``die linke" from the AlgorithmWatch data are compared to the search suggestions for the queries ``grüne'' and ``dielinke''.} For both data-sets we consider queries submitted between August 4st and September 30th, 2017. We programmed a web client which made calls to Google's query suggestion API \footnote{\url{http://clients1.google.de/complete/search}}. The crawling ran usually twice a day around 05:00 and 17:00 CET and produced an outcome of $401$ unique suggestion terms during the observation period. A sample of the data set can be seen in table \ref{tab:dataset}. The data is published under a Creative Commons licence \cite{bonart2018}. 

\begin{table}[ht]
\centering
\begin{tabular}{ccccc}
  \toprule
 source	& queryterm	& date	& suggestterm	& position\\
 \midrule
google & Alexander Gauland& 2017-08-04 05:30:51 & twitter & 0\\
google & Alexander Gauland& 2017-08-04 05:30:51 & itate & 1\\
google & Alexander Gauland& 2017-08-04 05:30:51 & kontakt & 2\\
google & Alexander Gauland& 2017-08-04 05:30:51 & dorothea gauland & 3\\
google & Alexander Gauland& 2017-08-04 05:30:51 & boateng & 4\\
google & Alexander Gauland& 2017-08-04 05:30:51 & krawatte & 5\\
google & Alexander Gauland& 2017-08-04 05:30:51 & carola hein & 6\\
google & Alexander Gauland& 2017-08-04 05:30:51 & ehefrau & 7\\
google & Alexander Gauland& 2017-08-04 05:30:51 & youtube & 8\\
google & Alexander Gauland& 2017-08-04T05:30:51 & islam & 9\\
$\cdots$ & $\cdots$ & $\cdots$ & $\cdots$ & $\cdots$ \\
\bottomrule
\end{tabular}
\caption{Illustration of the query suggestions data-set.}
\label{tab:dataset}
\end{table}

For the AlgorithmWatch data we restrict the analysis to organic search results for requests submitted from within Germany using a German keyboard layout. We also cleaned the data from errors as described in \cite{zweig2018}. This results in \numprint{896333} unique result lists from \numprint{1441430} single requests.

While the comparison of different rankings from query suggestions is straightforward this is not true for the search results. Although there is only a small effect of search personalization in the data the ranking usually slightly differs between search engine users. However, the public data-set does not provide any persistent user/plugin ids which prevents the comparison of result lists on an individual level over time. As a workaround, for each time-point we construct an aggregated ``typical'' search result list. This is done by calculating the average position of each URL which is present in more than a third of the requests for this time-point and query. 

\section{First insights and discussion}

Typically, result and suggestion lists of search engines are incomplete, top-weighted and indefinite. The importance of the items decreases with the depth which justifies a truncation of the list. As a measure of stability we therefore rely on the so called rank-biased overlap (RBO) statistic. It measures the similarity of two indefinite possibly non-conjoint lists and properly incorporates the characteristics of indefinite rankings \cite{webber2010}.  

\begin{figure}[p]
 \figuretitle{Stability of search results and suggestions over time: Successive comparison}
\centering
\includegraphics[width=\textwidth]{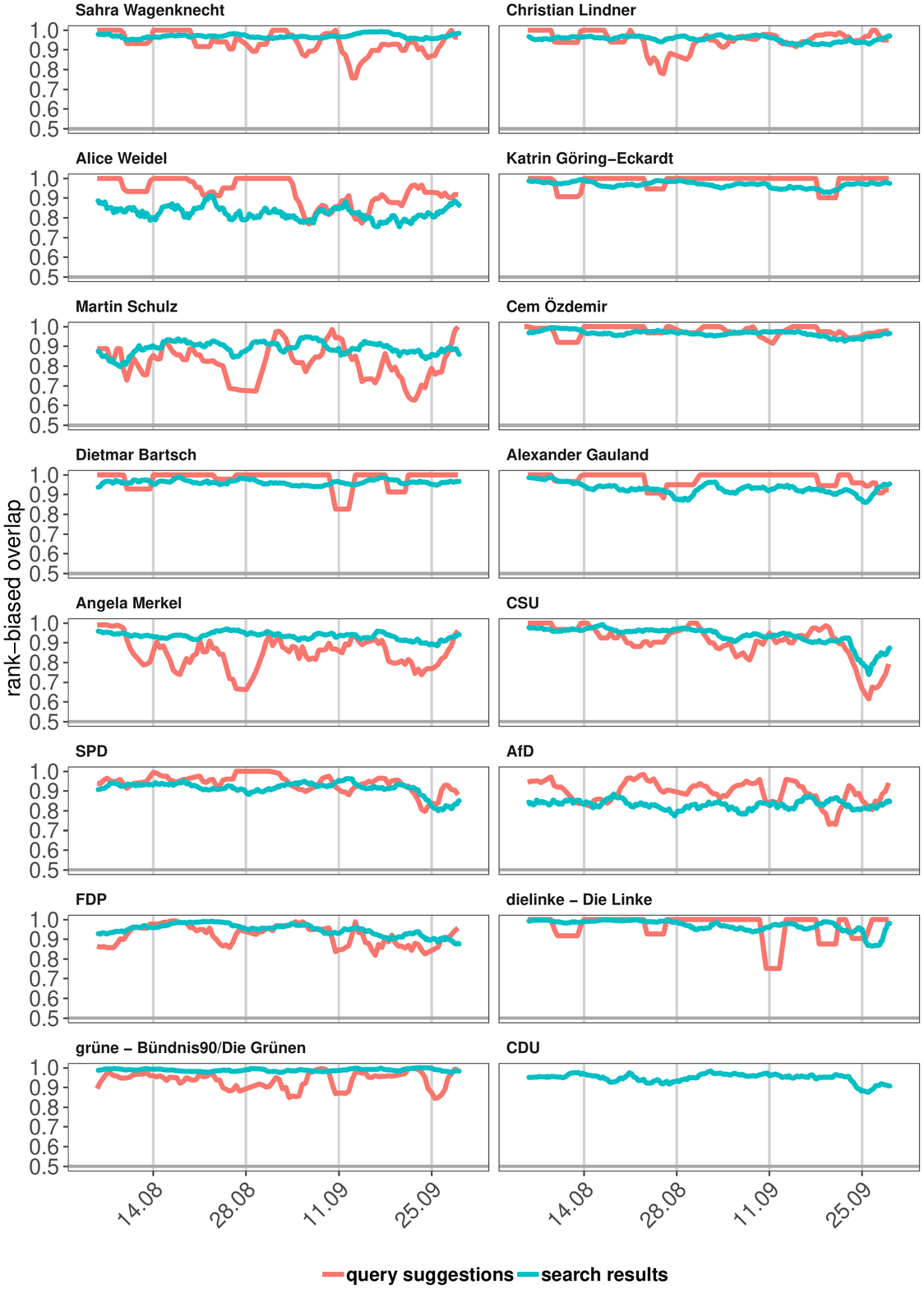}
\caption{For each keyword the search results (6 per day) and suggestions (2 per day) from each time-point are compared to the result from the \emph{previous} time-point. The rank biased overlap measure with $p = 0.85$ is used. A moving average with a window size of $n = 18$ and $n = 6$ (e.g. 3 days) is applied to smooth the time series. The horizontal gray line marks a RBO of $0.5$. \footnotesize{*see footnote \ref{fn:missing}.}}
\label{fig:rbo1}
\end{figure}

\begin{figure}[p]
 \figuretitle{Stability of search results and suggestions over time: Fixed comparison}
\centering
\includegraphics[width=\textwidth]{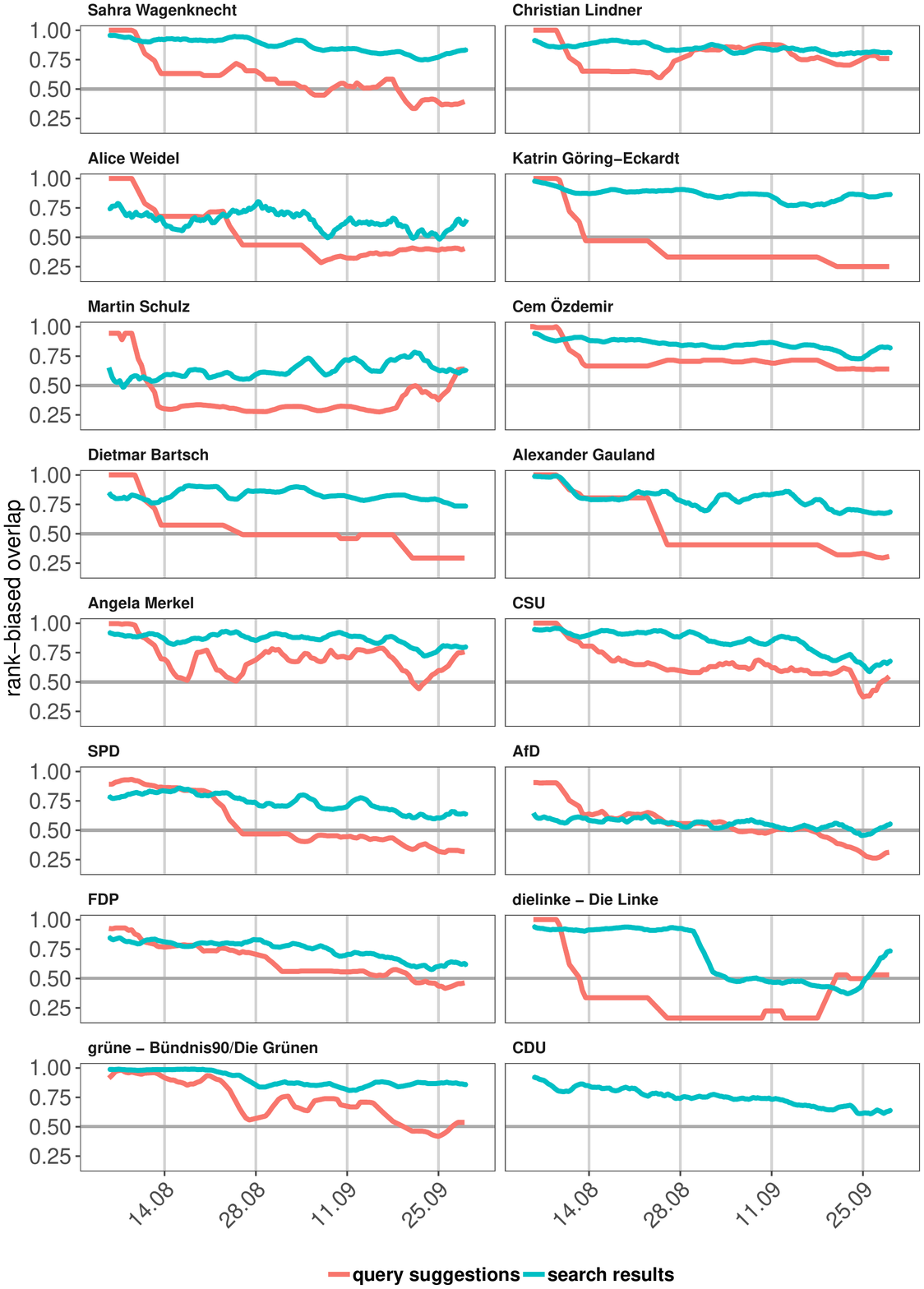}
\caption{For each keyword the search results and suggestions from each time-point are compared to the result from the \emph{earliest} time-point. The RBO measure with $p = 0.85$ is used. A moving average with a window size of 3 days is applied to smooth the time series. The horizontal gray line marks a RBO of $0.5$. \footnotesize{*see footnote \ref{fn:missing}.}}
\label{fig:rbo2}
\end{figure}

The RBO statistic can take values between $0$ indicating no similarity and $1$ for similar lists. The degree of top-weightiness can be controlled by a parameter $p$. For $p \rightarrow 0$ the first item in both lists are assigned all the weight while for $p \rightarrow 1$ all list items become equally important. For the analysis we choose $p = 0.85$ as this puts almost all the weight ($93\%$) on the first $10$ list items and results in an expected evaluation depth of $6.7$ list items. For each query and time-point we compare the ordered lists of aggregated search results and query suggestions with the preceding list (figure \ref{fig:rbo1}) and with the list from the earliest time-point in the sample (figure \ref{fig:rbo2}). To smooth the time series, a moving average with a window of three days is applied to the RBO values.

The figures show that the degree of stability for the query suggestions and the result sets mainly follow broad similar trends. Obviously, the composition of the lists changes during the two month observation period but the stability seems relatively high: Comparing the last result list with the first one yields a minimum RBO of around $0.5$ for most queries. For some queries, the query suggestions seem to be less stable.

This initial analysis only gives a broad overview over the general trend and can shed light to some interesting patterns in the data (e.g. the negative peak in figure \ref{fig:rbo1} for both search results and query suggestions for the term ``CSU'' around the date of the federal election). To better understand the role of specific events and how search results and query suggestions influence each other a more detailed look is required. For this it is important to study the actual, possibly dynamic, \emph{semantic} similarities between the data-sets over time.  It may be beneficial to include the top news headlines which appear in the organic search result lists to capture the influence of specific important events on the composition of result and suggestion lists.

In a political context it can also be interesting to ask whether query suggestions do lead to new and more diverse information sources. It could be possible that search suggestions for a specific term mainly resemble the topics and trending events listed on the term's result page. If this is the case, the effect of query suggestions on a user's process of political opinion formation could be neglected regardless whether he or she makes use of this feature or not.   

\section*{Acknowledgments}
\label{sec:Acknowledgments}

This research was supported by the Digital Society research program funded by the Ministry of Culture and Science of the German State of North Rhine-Westphalia.

\newpage
\bibliographystyle{splncs03}
\bibliography{literature}

\end{document}